\documentclass[utf8,bibauthoryear]{aa}
\usepackage{graphicx}
\usepackage{txfonts}
\usepackage{siunitx}
\usepackage[utf8]{inputenc}
\usepackage{natbib}
\usepackage{caption}
\begin{document}

   \title{Rotational spectral modulation of cloudless atmospheres for L/T Brown Dwarfs and
     Extrasolar Giant Planets}

   \author{P. Tremblin \inst{1}\thanks{\email{pascal.tremblin@cea.fr}}
     \and M. W. Phillips  \inst{2}
     \and A. Emery     \inst{1}
     \and I. Baraffe   \inst{2,3}
     \and B. W. P. Lew \inst{4}
     \and D. Apai      \inst{5,6}
     \and B. A. Biller \inst{7,8}
     \and M. Bonnefoy  \inst{9}
   }

   \institute{Maison de la Simulation, CEA, CNRS, Univ. Paris-Sud, UVSQ,
     Universite\'e Paris-Saclay, F-91191 Gif-sur-Yvette, France\
     \and Astrophysics Group, University of Exeter, EX4 4QL Exeter, UK\
     \and Ecole Normale Superieure de Lyon, CRAL, UMR CNRS 5574, 69364, Lyon
     Cedex, France\
     \and Lunar and Planetary Laboratory, The University of Arizona, 1640
     E. University Blvd,Tucson, AZ 85718, USA\
     \and Department of Astronomy and Steward Observatory, The University of
     Arizona, 933 N. Cherry Avenue, Tucson, AZ 85721, USA\
     \and Lunar and Planetary Laboratory, The University of Arizona, 1640
     E. University Blvd., Tucson, AZ 85721, USA\
     \and  Institute for Astronomy, University of Edinburgh, Blackford Hill,
     Edinburgh EH9 3HJ, UK\
     \and Centre for Exoplanet Science, University of Edinburgh, Edinburgh, UK\
     \and Univ. Grenoble Alpes, CNRS, IPAG, 38000 Grenoble, France
   }

   \date{Received \#\#\# \#\#\, 2019; accepted \#\#\# \#\#, 2019}

  \abstract
   {}
   {The rotational spectral modulation (spectro-photometric variability) of brown dwarfs is
     usually interpreted as 
     a sign of the presence of inhomogeneous cloud covers in the
     atmosphere. This paper aims at exploring the role of temperature
     fluctuations in these spectral modulations.
   These fluctuations could naturally arise in a convective atmosphere impacted
   by diabatic processes such as complex chemistry, i.e. the recently proposed
   mechanism to explain the L/T transition: CO/CH$_4$ radiative convection.}
 {After exploring the observed spectral-flux ratios between different objects
   along the cooling sequence, we use the 1D radiative/convective code
   \texttt{ATMO} with ad-hoc modifications of the temperature gradient to
   model the rotational spectral modulation of 2MASS~1821, 
   2MASS~0136, and PSO~318.5-22. We also explore the impact of CH$_4$ abundance
   fluctuations on the spectral modulation of 2MASS~0136.}
   {The spectral-flux ratio of different objects along the cooling sequence and
     the rotational spectral modulation within individual objects at the L/T
     transition have similar characteristics. This strongly suggests that the
     main parameter varying along the cooling sequence, i.e. temperature,  might
     play a key role in the rotational spectral modulations at the L/T
     transition. Modeling the spectral bright-to-faint ratio of the modulation
     of 2MASS~1821, 2MASS~0136, and PSO~318.5-22 shows that
     most spectral characteristics can be reproduced by temperature
     variations alone. Furthermore, the approximately anti-correlated
     variability between different wavelengths can be easily
      interpreted as a change in the temperature gradient in the
     atmosphere which is the consequence we expect from CO/CH$_4$
     radiative convection to explain the L/T transition. The deviation from an
     exact anti-correlation could then be interpreted as a phase shift similar
     to the hot-spot shift a different bandpasses in the atmosphere of hot Jupiters.}
   {Our results suggest that the rotational spectral modulation from cloud-opacity
     and temperature variations are degenerate. If the nearly
     anti-correlated signal
     between different wavelengths is indeed a strong sign of a change in the
     temperature gradient, the detection of direct cloud spectral signatures,
     e.g. the silicate absorption feature at 
     10 $\mu$m, would help to confirm the presence of clouds and their contribution
     to spectral modulations (which does not exclude temperature variations or
     other mechanisms to be also at play). Future studies
     looking at the differences in the spectral modulation of objects with and
     without the silicate absorption feature may give us some insight on how to
     distinguish cloud-opacity fluctuations from temperature fluctuations. }


   \maketitle
%

\section{Introduction} \label{sec:introduction}
One of the key feature of brown dwarfs at the L/T transition is the rotational
spectral modulation (spectroscopic variability) of the object \citep[see][for a
  review]{biller:2017,artigau:2018}. Spectral modulations are of great interest
since they open the possibility to probe spatial variations in the vertical
structure of the atmospheric columns, as different wavelengths probe different
pressure levels. Up to now,
most of the spectroscopic modulations have been observed using the
Wide Field Camera 3 aboard the {\it Hubble Space Telescope} (HST)
\citep[e.g.,][]{buenzli:2012,apai:2013}, and a wealth of new data will be brought
to the community by the {\it James Webb Space Telescope} (JWST). Among the
principal characteristics of the rotational spectral modulations observed by HST
is the presence 
of increased variability amplitude within the 1.45 $\mu$m water absorption feature
and, for L-type dwarfs, the apparent absence of differential modulations in- and
out of the water band \citep{yang:2015}.
Simultaneous HST and Spitzer time-resolved observations found prominent phase
shifts between lightcurves observed at different wavelength (i.e., pressure
levels). \citet{buenzli:2012} found a pressure-dependent phase shift pattern in
the late-T dwarf 2M2228, a pattern that was found again, years later
\citep{yang:2016}. Similarly, \citet{biller:2018} detected very large phase
shifts in the planetary-mass brown dwarf PSO~318.5-22. \citet{yang:2016}
reported phase shifts 
in some brown dwarfs (with values ranging from 30 to 180 degrees). In contrast,
\citet{apai:2013} showed that no phase shifts are present in two
L/T transition brown dwarfs. These patterns are not yet understood, but the fact
that they appear to be stable for a given object over hundreds of rotational
periods suggests that the large-scale horizontal-vertical structures they trace
are characteristic to the objects and, thus, may hold important clues to the
nature of the spatial-spectral differences observed \citep[although Luhman 16B
 can show really significant changes from night to night, e.g.][]{gillon:2013}.
 Medium-resolution spectroscopy (at spectral resolving power R~4,000) has also 
been used to study spectral variability on Luhman 16AB exploring
variability at the resolution of spectral lines 
\citep[e.g. KI absorption feature][]{faherty:2014,kellogg:2017}. High-resolution
spectroscopy enabled Doppler imaging 
to create possible maps of the the brightness variations of Luhman 16B
\citep{crossfield:2014}.

The observed photometric and spectral modulations are
often interpreted as due to the presence of an inhomogeneous cloud cover in the
atmosphere \citep[e.g.][]{marley:2010}. While highest-amplitude (J-band)
modulations are found at the L/T transition for old field objects
\citep{radigan:2014} and should 
probe deep clouds in the
atmosphere of old, high surface gravity objects, a physical understanding of why
the spatial structure of clouds should change at the L/T transition is still
currently lacking.
\citet{robinson:2014} has also proposed that temperature fluctuations can also
 be responsible for spectroscopic
 variability and could potentially explain, in part, the pressure-dependent
 phase shifts observed in 2M2228 by \citet{buenzli:2012}. However,
 \citet{radigan:2012,apai:2013} looked into effective 
 temperature  variations only (i.e., by modeling the spectral modulations via
 linear combinations of two one-dimensional atmosphere models that differed in
 effective temperature, but not in the cloud prescription) and have found that
 it  cannot explain the changes in broadband, near-infrared colors.

 \citet{tremblin:2015,tremblin:2016,tremblin:2017} have shown 
 that the spectra and luminosity of L/T transition brown dwarfs, along with the
 reddest low-gravity objects, can be reproduced by cloudless models with a
 reduced temperature gradient. So far, spectroscopic variability arising from a
 change in the temperature gradient of the atmosphere (which is different from a
 change of effective temperature) has never been explored in the literature. 
 \citet{tremblin:2019} has shown that such a
 reduced gradient should 
 indeed be expected when convection is impacted by thermal and compositional
 diabatic processes: diabatic convection leads to thermohaline and moist
 convection in Earth's atmosphere and oceans, reducing their temperature
 gradient. In brown dwarfs, CO/CH$_4$ (and N$_2$/NH$_3$) radiative convection is a
 direct analog and should also reduce the temperature gradient of the
 atmosphere. When CO in the atmosphere is converted into CH$_4$ the diabatic
 processes stop and the convective system switches between the diabatic
 and adiabatic convective behavior \citep[a so-called ``cooling crisis'' by analogy to
 the boiling crisis in two-phase convective flows][]{tremblin:2019}.This could
 naturally explain the L/T transition with an increase of the temperature that
 could also naturally reproduce the J-band brightening and FeH resurgence at the
 transition. Since we expect a change in the temperature gradient of the
 atmosphere at the L/T transition in this scenario, it is natural to expect
 rotational spectral modulations to arise from spatial thermal inhomogeneities in
 the atmosphere.

 In this paper, we explore the possibility of reproducing representative
 examples of observed spectral modulations by introducing temperature variations
 in the atmosphere. We first 
 study the HST spectral evolution of different objects along the cooling
 sequence of brown dwarfs and describe the model used for the spectral modeling
 (Sect.~\ref{sec:model}). Then we model the HST rotational spectral modulation of the L
 dwarf 2MASS~J18212815+1414010 (2MASS~1821 hereafter) and the T dwarf
 2MASS~J01365662+0933473 (2MASS~0136 hereafter) (Sect.~\ref{sec:ltdwarfs}), and
 the HST and Spitzer variability of PSO~318.5-22 (Sect.~\ref{sec:pso}). We
 finally discuss the impact  of the modulations on the JWST spectral coverage
 (Sect.~\ref{sec:jwst}) and the impact of CH$_4$ variations in the T-dwarf regime
 (Sect.~\ref{sec:ch4}) before reaching our conclusions and discussion
 (Sect.~\ref{sec:conclusion}).

 \begin{figure}[btp]
\begin{centering}
\includegraphics[width=1.0\columnwidth]{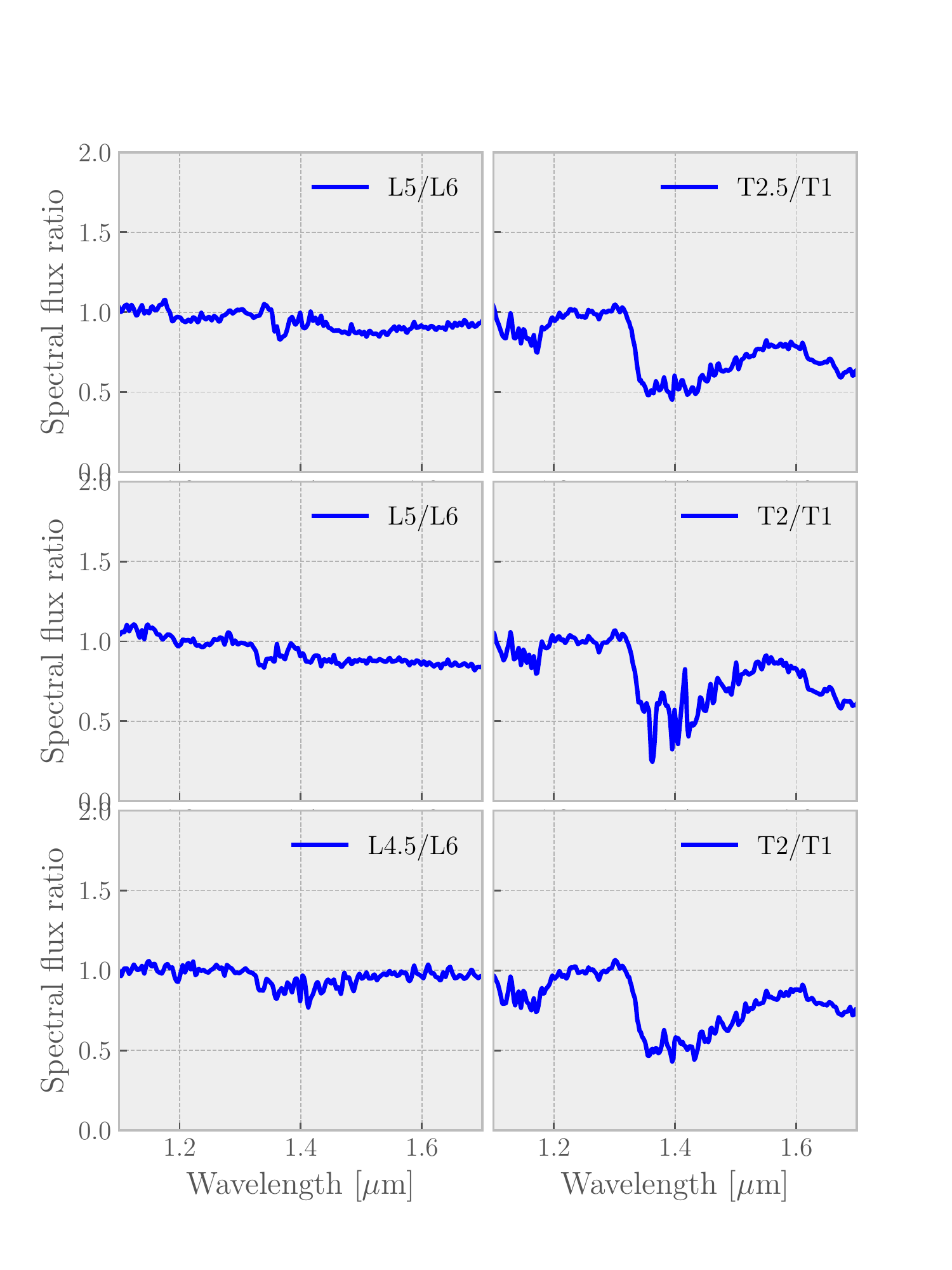}
\caption{Left: spectral-flux ratio of the L5 dwarf 2MASS J01550354+0950003, L5
  2MASSW J1507476-162738, L4.5 SDSS J083506.16+195304.4 to a reference L6 dwarf
  2MASSI J1010148-040649. Right: spectral-flux ratio of the T2.5 dwarf IPMS
  J013656.57+093347.3, T2 2MASS J11220826-3512363, T2 SDSSp J125453.90-012247.4,
to a reference T1 dwarf SDSS J085834.42+325627.7. The data are SpeX prism
spectra coming from the SpeX Prism Library
\citep{burgasser:2014}  and are similar to spectral standards. } \label{fig:ltdwarfs}
\end{centering}
 \end{figure}

\section{Spectral evolution along the cooling sequence and model
  description} \label{sec:model}

\subsection{Spectral evolution along the cooling sequence of brown dwarfs}

We first look at the spectral evolution of objects of different
spectral types along the cooling sequence. We assume that they do not show
internal spectroscopic modulations but look at the relative change between different
spectral types at different stages in the cooling sequence.
Figure~\ref{fig:ltdwarfs} shows the
observed spectral-flux ratios in WFC3/IR/G141 bandpass 
for different objects along the cooling sequence, at the L/T transition, for
late L dwarfs (ratio of L4.5/L5 to a L6) and early T dwarfs (ratio of T2/T2.5 to
a T1). The spectral ratio for L dwarfs is essentially flat with no
spectral signature. On the contrary, a strong signature can be identified in the
spectral ratio for T dwarfs 
with a suppressed absorption amplitude in the 1.4-$\mu$m water band relative to the
adjacent continuum. Therefore, the evolution of the relative water amplitude 
as a function of spectral type along the cooling sequence is similar to the
spectral modulations of this feature in individual mid-L and T brown dwarfs
\citep[e.g.][]{yang:2015}. This similarity between the spectral evolution along the
cooling sequence and the rotational spectral modulations within individual
objects has also been pointed out by \citet{kellogg:2017}.

Given the similarity, 
we can expect that the main parameter changing along the cooling sequence
(i.e., temperature) might also play a significant role in the rotational spectral modulations
of individual objects.  Thus, this similarity suggests that spatial variations
may exist in the pressure-temperature profiles within individual brown
dwarfs. Although \citet{radigan:2012} and \citet{apai:2013}, among others, have
shown that  
variations in the effective temperature alone cannot reproduce the observed
rotational color modulations, it is likely that a combination of the
parameters evolving along the cooling sequence and including the effective
temperature can reproduce the spectral modulations.
The recent introduction of temperature
gradient modifications \citep{tremblin:2015} motivates the exploration of
temperature-gradient fluctuations as such a
possible source of spectral modulations in the atmospheres of brown dwarfs.

\subsection{Model description}

Although spatial inhomogeneities leading to rotational modulations can be long-lived
stationary patterns, it is likely that, in general, such inhomogeneities are
associated with a time-dependent process on some finite timescale. Strictly
speaking, this implies that one should model these patterns with a
multi-dimensional, time-dependent simulation that is capable of capturing the appropriate
physics on the adequate length- and timescales (e.g., hydrodynamics including
convection, chemical reactions, radiative transfer). Strictly speaking,
stationary 1D codes are not appropriate for such studies since spatial
inhomogeneities and time variability violate the homogeneous and stationary
assumptions. Nonetheless, 3D time-dependent tools are very
challenging to build (if not impossible depending on the needed length- and
timescales) and in this paper, we adopt a much simpler approach, based on 1D
models. This simpler and more efficient approach, however, comes with some caveats.

We use the 1D atmospheric code \texttt{ATMO} that has been previouly used to
model the spectra of a broad range of different brown dwarfs
\citep{tremblin:2015,tremblin:2016,tremblin:2017, phillips:2020}. The radiative transfer scheme
is described in \citet{amundsen:2014,amundsen:2016} and the chemistry scheme in
\citet{drummond:2016} and \citet{goyal:2018} for rainout. The chemistry includes
279 species for gas phase chemistry, ions chemistry, and condensible
species. The radiative transfer uses 22 opacity sources (H$_2$-H$_2$ CIA,
H$_2$-He CIA, H$_2$O, CO, CO$_2$, CH$_4$, NH$_3$, Na, K, L, Rb, Cs, TiO, VO,
FeH, PH$_3$, H$_2$S, HCN, C$_2$H$_2$, SO$_2$, Fe, H$^-$ free-free and
bound-free) including also Rayleigh scattering.

Depending on global
parameters such, as effective temperature $T_\mathrm{eff} $ and surface
gravity log(g), the code finds a time-independent pressure/temperature (PT)
structure that satisfies hydrostatic equilibrium and energy conservation either
with radiative energy transport or convective energy
transport (often referred as 1D self-consistent model). \citet{tremblin:2016}
has found that the observed low near-infrared 
fluxes can be very well reproduced if the temperature in the atmosphere is lower
than what is predicted by radiative/convective equilibrium-based model. This effect is a
possible natural consequence of diabatic convection in the form of CO/CH$_4$
radiative convection in the brown dwarf context. This is mimicked in \texttt{ATMO}
by a reduction of the adiabatic index to an effective index
$\gamma_\mathrm{eff}$ in between two pressure levels $P_{\gamma,\mathrm{min}}$ and
$P_{\gamma,\mathrm{max}}$ that reduces the adiabatic gradient followed by convection in a
given region.

For clarity, we list below the
different approximations to keep in mind when using such a code to study
rotational modulations:
\begin{itemize}
    \setlength\itemsep{1em}
\item Since the code is stationary and 1D,
  a first approximation is to model the bright and faint states of the
  spectral modulation with a combination of 1D PT profiles. Since the code only
  solves for the hydrostatic balance in 1D, we can only model local ad-hoc
  temperature and compositional variations and we cannot self-consistently model
  3D dynamical processes like convection and winds and explore directly the
  lightcurve time evolution. We therefore focus our study on modeling the
  spectral characteristics of the bright-to-faint ratio of the spectral
  modulations. 

\item Since the code finds a solution for some global parameters, we will modify
  the PT profile in an ad-hoc way to mimick a bright or faint modulation
  (Sect.~\ref{sec:ltdwarfs}) or modify directly the global parameters to model the
  hemispherically averaged bright state (Sect.~\ref{sec:pso}).
  The exact value of these parameters is therefore not really meaningful (also
  probably degenerate) and only the resulting difference in the PT profile in
  the regions that do impact the observed spectrum really makes sense.

\item In \texttt{ATMO}, chemistry is solved at equilibrium or is coupled to a
  full chemical network \citep{venot:2012}, finding the stationary solution of
  the chemical kinetic equations on a very long timescale (typically 10$^{12}$~s) with
vertical quenching \citep[see for details][]{tremblin:2015}. Since we do not
know a-priori the impact of horizontal 
quenching caused by winds and convection, we do not know which chemical species
is at chemical equilibrium or horizontally-quenched by the dynamics. Therefore, for
simplicity, we assume the two limit cases: chemical abundances are at equilibrium in the
modulation or are kept constant to the abundances of the background PT
profile. Furthermore we simplify 
vertically-quenched out-of-equilibrium chemistry by assuming a fixed
and constant abundance profile of CH$_4$ needed to reproduce the
spectrum. In Sect.~\ref{sec:ltdwarfs} and Sect.~\ref{sec:pso}, the CH$_4$
abundance is therefore kept constant and we explore the variability associated
to CH$_4$ abundance fluctuations in Sect.~\ref{sec:ch4}.
\end{itemize}

\begin{figure}[btp]
\begin{centering}
\includegraphics[width=1.0\columnwidth]{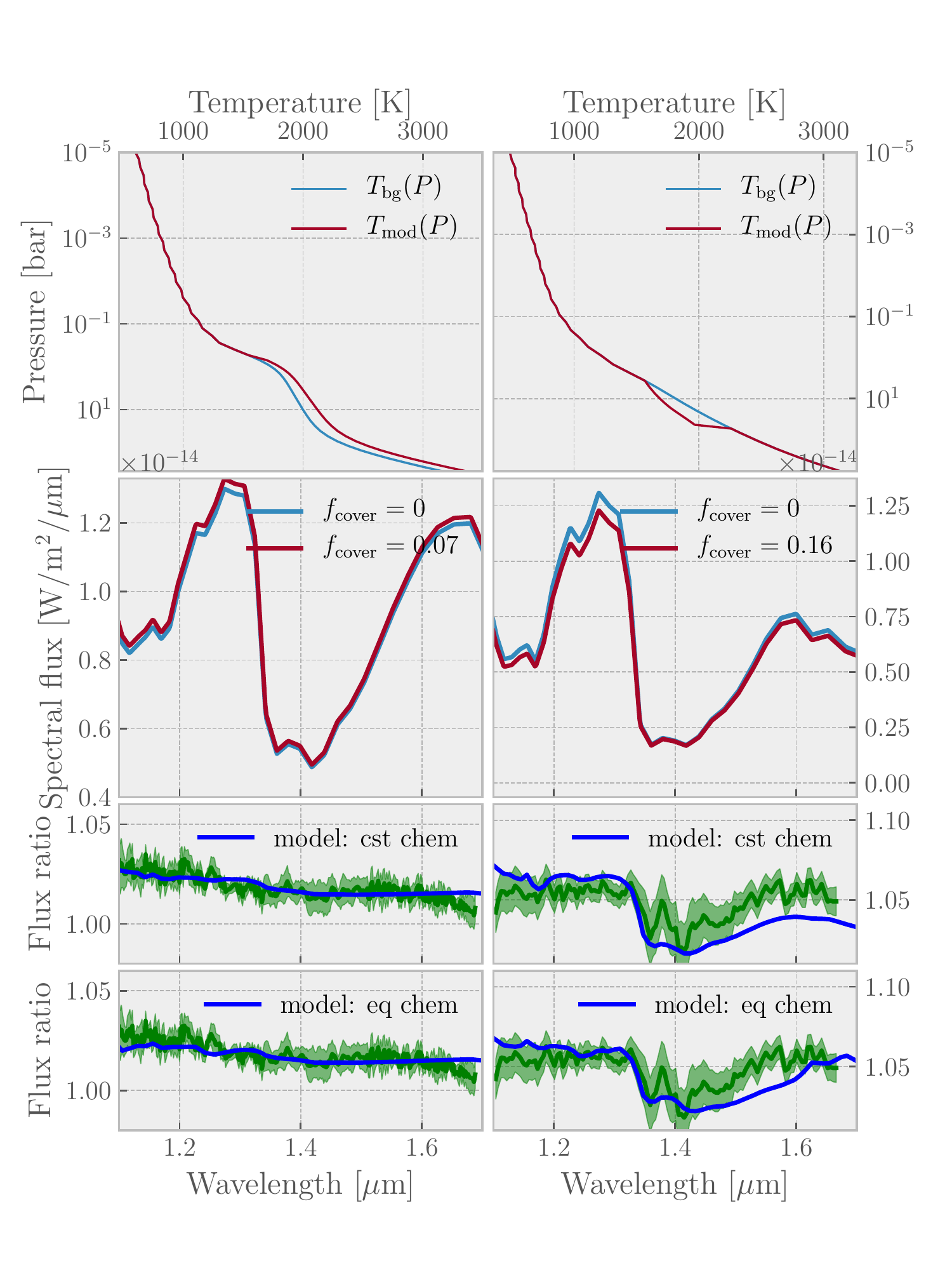}
\caption{Left: 2M~1821 (L dwarf), right: 2M~0136 (T dwarf). Top: PT profiles,
  blue is the background 
  profile and red is the modulation PT profile. Middle: Spectral flux at a resolution
  R$\sim$100, blue is the spectral flux computed with the background PT profile,
  red is the spectral flux computed with Eq.~\ref{eq:flux}. Bottom: Modeled
  spectral bright-to-faint ratio in blue assuming either equilibrium chemistry
  in the modulation or constant abundances equal to the background values. Observed
  spectral ratio is in green \citep{yang:2015}.} \label{fig:2M1821-2M0136}
\end{centering}
\end{figure}

\begin{table}
\centering
\begin{tabular}{l|l|l}
  & 2M~1821 & 2M~0136 \\
  \hline
  \hline
  T$_\mathrm{eff}$      & 1650          & 1400 \\
  logg                 & 4.5   &  5\\
  $\gamma_\mathrm{eff}$ & 1.05          & 1.15 \\
  $P_{\gamma,\mathrm{min}}$ [bar]& 6.3$\times$10$^{-1}$ &1.5  \\
  $P_{\gamma,\mathrm{max}}$ [bar]& 6.3$\times$10$^{1}$ & 5$\times$10$^{2}$\\
  $$[M/H]              & 0     & 0 \\
  X(CH$_4$)            & $10^{-10}$ & 8$\times$10$^{-5}$\\
  \hline
  $P_{\mathrm{mod,min}}$ [bar]& 7.1$\times$10$^{-1}$ &3.7  \\
  $P_{\mathrm{mod,max}}$ [bar]& 2.8$\times$10$^{2}$ &5.4$\times$10$^{1}$  \\
  $T_\mathrm{A}$ [K] & $T_\mathrm{bg}(P)+60$ & $T_\mathrm{bg}(P_{\mathrm{mod,min}})$\\
  $T_\mathrm{B}$ [K] & 138 & 440\\
  $f_\mathrm{cover}$ & 0.07 & 0.16 \\
\end{tabular}
\caption{Parameters used to model the background and modulation PT profiles of 2M~1821 and
  2M~0136.}\label{tab:2M1821-2M0136}
\end{table}

\section{Results on spectral variability modeling} \label{sec:results}

\subsection{Variability of 2MASS~1821 and 2MASS~0136}\label{sec:ltdwarfs}
As a first application, we study the spectral modulations observed by HST in the L5 dwarf
2M~1821 and in the T2 dwarf 2M~0136 \citep{yang:2015}. For these objects, we
first compute a background model with a PT profile given by $T_\mathrm{bg}(P)$
and a parametrization of the PT profile of a modulation, given by
Eq.~\ref{eq:mod}. 
\begin{equation}\label{eq:mod}
T_\mathrm{mod}(P) = T_\mathrm{A}(P) + \frac{(\log\tau(P) -
  \log\tau(P_\mathrm{mod,min}))}{(\log\tau(P_\mathrm{mod,max}) -
  \log\tau(P_\mathrm{mod,min}))} T_\mathrm{B}
\end{equation}
with $\tau(P)$ the optical depth profile at a wavelength of 1.25 $\mu$m. We then
 compute the total flux by 
\begin{equation}\label{eq:flux}
F_\mathrm{tot} = f_\mathrm{cover} F_\mathrm{mod} + (1-f_\mathrm{cover})
F_\mathrm{bg}
  \end{equation}
with $F_\mathrm{bg}$ and $F_\mathrm{mod}$ the surface fluxes computed with the
background and modulation PT profile, respectively. In
Tab.~\ref{tab:2M1821-2M0136}, we summarize the parameters that we have used to model the
bright and faint states of these two dwarfs. The abundance of CH$_4$ used for
2M~1821 is low because CH$_4$ is not sufficiently abundant in the L-dwarf regime
to be observable yet. In the T-dwarf regime, for 2M~0136, CH$_4$ begins to be
observable. We keep its abundance constant in this section but explore the
effect of spatial variations in its abundance in a dedicated section (Sect.~\ref{sec:ch4}).
in Fig.~\ref{fig:2M1821-2M0136}, we show the resulting PT profiles, spectra,
modeled spectral ratio, compare it to the observed spectral ratio from
\citet{yang:2015}. We find that a
covering fraction between 7~\% and 16~\% is needed to reproduce the 
observations, however, the covering fraction is highly degenerate with the
magnitude of the modulations in the PT profiles. Both models seem
to reproduce well the 1.1-1.5~$\mu$m part of 
the spectrum, but deviate in the 1.5-1.7 $\mu$m region. Nevertheless, we find that the main
characteristics i.e. the presence/absence of increased variability amplitude
within the 1.45 $\mu$m water absorption feature, are
well reproduced. Overall, assuming constant abundances from the background PT
profile or chemical equilibrium in
the modulation profile does not seem to significantly change the flux ratio in
the HST WFC3 bandpass (see Sect.~\ref{sec:jwst} and \ref{sec:ch4} for the impact
on the JWST spectral coverage). Equilibrium
chemistry seems to produce a slightly better agreement with the data.
This level of agreement is relatively good since the parameter is explored
manually, a better match could potentially be achieved by automatic
algorithms and bayesian analysis. However, we emphasize again that these modifications are
probably degenerate and another combination could lead to a similar result. We
point out that 2M~0136 is likely a member of the Carina-Near moving group
\citep{gagne:2017} whith an estimated age of 200 Myr. Looking at evolutionary
models between 120 and 500 Myr \citep{phillips:2020}, we expect a surface gravity
of  $\log(g)=$~4.5-5.0 for the
effective temperature of 2M~0136, which is consistent with the surface gravity
we have used in the model. Despite being a moving group member, 2M~0136 is not necessary a
low-surface gravity object, hence we do not expect our results to be affected by
its young age. 

\begin{table}
\centering
\begin{tabular}{l|l|l}
  & \multicolumn{2}{c}{PSO~318.5-22}   \\
  \hline
  \hline
  & bright & faint  \\
  \hline
  T$_\mathrm{eff}$      & 1275   & 1275\\
  $\gamma_\mathrm{eff}$ & 1.028   & 1.03 \\
  P$_\mathrm{\gamma,min}$ [bar]& 7$\times$10$^{-2}$ & 6$\times$10$^{-2}$ \\
  \hline
  P$_\mathrm{\gamma,max}$ [bar]& \multicolumn{2}{c}{10}\\
  logg                 & \multicolumn{2}{c}{3.7}\\
  $$[M/H]                & \multicolumn{2}{c}{0.4}\\
  X(CH$_4$) & \multicolumn{2}{c}{$10^{-10}$}
\end{tabular}
\caption{Parameters used to model the bright and faint state of
  PSO~318.5-22.}\label{tab:pso}
\end{table}

\begin{figure}[btp]
\begin{centering}
\includegraphics[width=1.0\columnwidth]{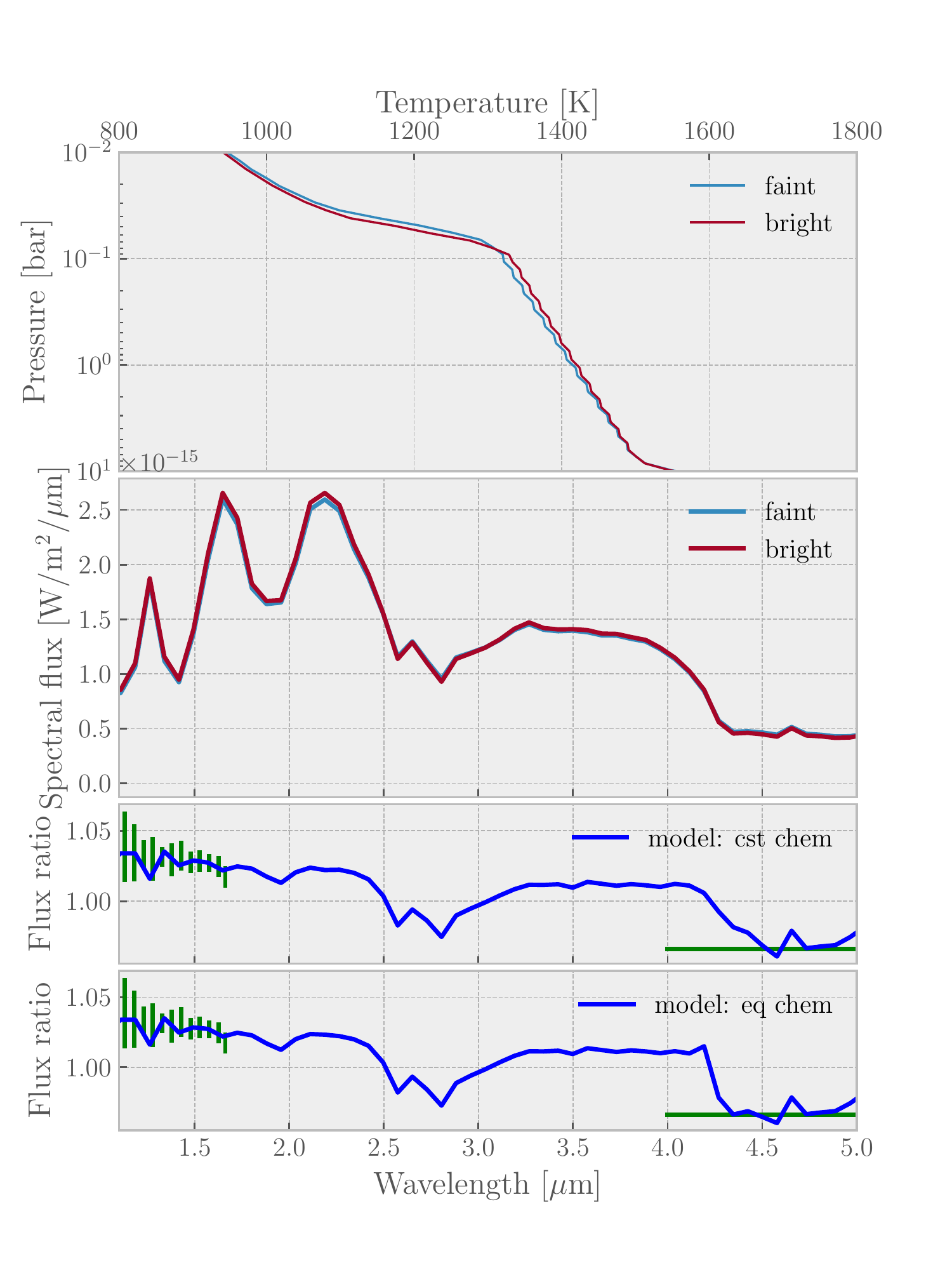}
\caption{Model of PSO~318.5-22. Top: PT structures, red is the bright
  state, blue is the faint state. Middle: Spectral flux at a resolution
  R$\sim$25, red is the bright state, blue is the faint state. Bottom: Modeled
  spectral ratio in blue either assuming chemical abundances in the bright state
  equal to the abundances in the faint state or assuming chemical
  equlibrium. The observed spectral ratio is in green
  \citep[see][, for details]{biller:2018}.} \label{fig:pso} 
\end{centering}
\end{figure}

\subsection{Variability of PSO~318.5-22}\label{sec:pso}

We now turn to the spectroscopic variability of PSO~318.5-22 and its observed
$\sim$200$^\circ$ phase shift between near- and mid-infrared signals. The simultaneous
HST-WFC3 (G141 grism) and Spitzer-IRAC (channel 2, central wavelength 4.5
$\mu$m) modulations of the  
free-floating planetary-mass object PSO~318.5-22 has been recently observed by
\citet{biller:2018}. These HST observations were five orbits long and catch a
lightcurve maximum in orbit 3 and get
close to two minima with different modulations in orbit 1 and 5.
The strong reddening and absence of strong CH$_4$ signature in the spectrum of
PSO~318.5-22 suggests that this object falls into the L-dwarf regime
\citep{tremblin:2017}. This seems to be confirmed by the absence of increased variability
in the 1.4-$\mu$m water feature for the orbit 3/orbit 5 ratio in \citet{biller:2018}
(uncertainties about this absence remain given the 
error bars and the orbit  3/orbit 1 ratio). As such, the spectral signatures of CH$_4$
are essentially absent and we use similarly to 2M~1821 a low CH$_4$ abundance to
reproduce the main spectral characteristics (see Tab.~\ref{tab:pso}). The other
parameters, e.g., effective temperature, surface gravity, metallicity, are similar
to the model used in \citet{tremblin:2019}.

Given the uncertainties in the modulations observed in PSO 318.5-22, we have obtained a
good model match with a simple 1D hemispheric-averaged bright and faint state,
hence we do not use here a more complex model.
In Fig.~\ref{fig:pso}, we show the PT structures, spectra, and the modeled and observed
spectal ratios. The observational data are taken from \citet{biller:2018}, using
the orbit 3/orbit 5 ratio of the WCF3 observations and a ratio of
-3.4$\pm$0.1\% in Spitzer IRAC channel~2, under the approximation that HST and
Spitzer data are nearly anti-correlated. The modeled spectral ratio shows that
we can reproduce relatively well the spectral modulations in the WCF3 bandpass
and, at the same time, the anti-correlated variability in IRAC Channel 2. This
mechanism can be described easily: since HST and Spitzer data are probing
different pressure levels, an increase in the HST flux and a decrease
in the Spitzer flux can be interpreted as an increase of temperature in the
layers probed by HST and a decrease of temperature in the layers probed by
Spitzer. This can be interpreted as a change in the temperature gradient between
the two pressure levels. Indeed, the PT structures in Fig.~\ref{fig:pso} shows
that the temperature for pressures lower than 10$^{-1}$ bar is decreasing going
from faint to bright whereas the temperature is increasing for pressures higher
than 10$^{-1}$ bar. Such a change in the temperature gradient has been proposed
to be at the origin of the L/T transition in \citet{tremblin:2016} and has been
shown to be a natural consequence of CO/CH$_4$ radiative convection in
\citet{tremblin:2019}. We will discuss in Sect.~\ref{sec:conclusion} the
possible implications of the departure from an exact anti-correlation between
HST and Spitzer data since we cannot address this point with a simplistic 1D
approach. 

\subsection{Extrapolation for the JWST spectral coverage}\label{sec:jwst}

\begin{figure*}[btp]
  \begin{centering}
    \includegraphics[width=1.0\linewidth]{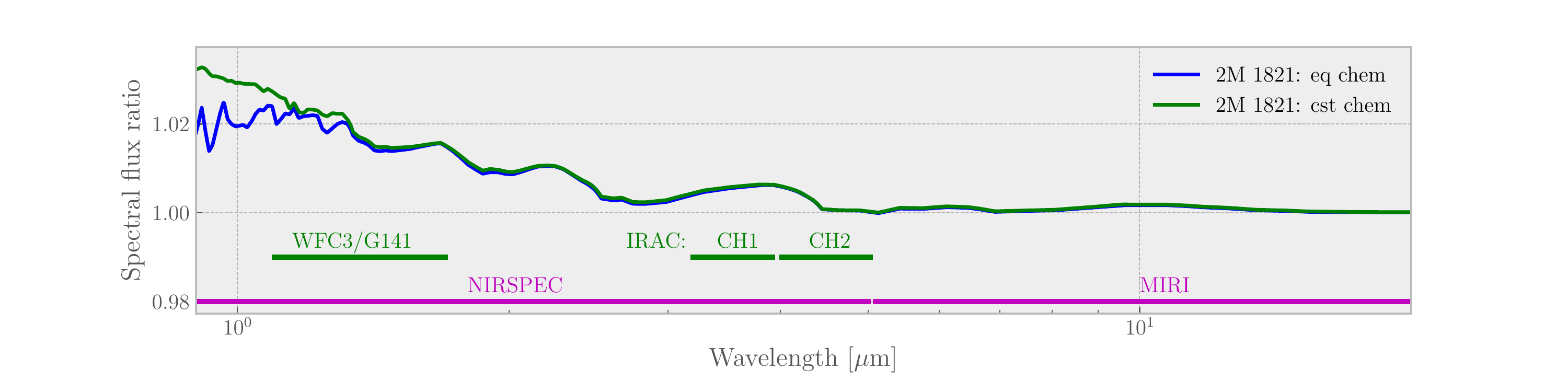}
    \includegraphics[width=1.0\linewidth]{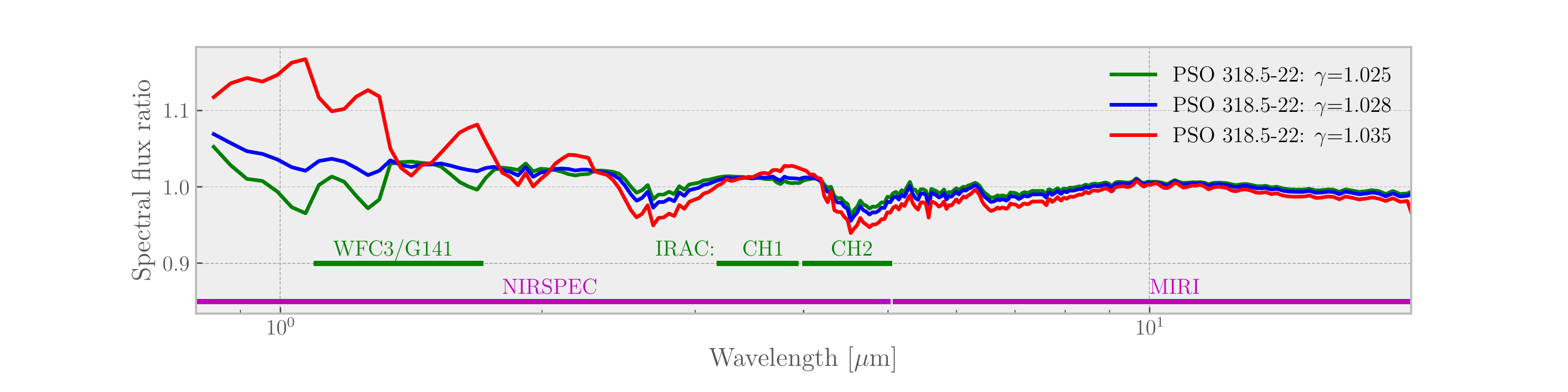}
\caption{ Top: Models of the bright-to-faint spectral ratio of 2M 1821 assuming
  constant or equilibrium chemistry in the modulation between 0.9 and 20
  $\mu$m. Bottom: Models of the bright-to-faint spectral ratio of PSO 318.5-22
  assuming different $\gamma$ values in the bright phase.} \label{fig:jwst}
\end{centering}
\end{figure*}

We show in the top panel of Fig.~\ref{fig:jwst}, the models for the spectral
ratio of 2M 1821  assuming equilibrium chemistry or constant chemistry in the
modulation between 0.9 and 20 $\mu$m. While the differences between the two types
of model are small in the HST WFC3 bandpass, they are relatively large in the
spectral window that can be explored with NIRSPEC at short wavelengths. The
bright-to-faint spectral 
ratio of 2M 1821 is smaller for equilibrium chemistry at wavelengths smaller
than 1.1 $\mu$m that can be 
explored with NIRSPEC. The smaller spectral ratio is because of the larger
difference in FeH abundance between the bright and faint states in the
equilibrium chemistry model. NIRSPEC will therefore be able to probe the impact
of FeH on 
spectral modulations of brown dwarfs and give constrains on the timescale and
kinetics on which the molecule is able to form in the modulations. The
bright-to-faint spectral ratio of 2M 0136 is mainly impacted by CH$_4$ abundance
variations and will be investigated in a dedicated section
(Sect.~\ref{sec:ch4}) since the carbon chemistry is known to be
out-of-equilibrium in brown-dwarf atmospheres \citep{fegley:1996}. 

We show in the bottom panel of Fig.~\ref{fig:jwst}, the models for the spectral
ratio of PSO 318.5-22 with different $\gamma$ values (different temperature
gradient) in the bright state, between 0.9 and 20 $\mu$m. An increase in the
temperature gradient in the bright phase ($\gamma$= 1.035) results in a larger
spectral ratio in the NIRSPEC spectral coverage relative to the MIRI
bandpass, which enhances the difference in anti-correlated spectral ratios
between the near-IR and mid-IR. 
A decrease in the temperature gradient in the bright phase ($\gamma$=1.025),
however, can remove the anti-correlation between the window at 1.05-1.25 $\mu$m
and mid-IR. The difference in the spectral ratio between near-IR and mid-IR is
therefore quite sensitive to modulations in the tempetature gradient in the
atmopshere.

We point out that the models presented in this paper have been constrained to
reproduce the bright-to-faint spectral ratios in the HST WFC3 and Spitzer channel 2
bandpasses. The spectral ratios at other wavelengths are extrapolations and not
really predictions: observed deviations to these models might simply mean that
the models need to include temperature-gradient modulations e.g. higher up in the
atmosphere to reproduce longer wavelengths that generally probe lower
pressures. Nonetheless, it is interesting to note that our models have little
variability at long wavelengths around 10 $\mu$m. This window could therefore be
used to probe modulations that could arise from spatial inhomogeneities in the abundance of
silicates. It could, therefore, help to disentangle between spatial inhomogeneities of
the temperature gradient and spatial inhomogeneities of a silicate cloud layer. Additional
models that include clouds and reproduce the 10-$\mu$m silicate feature will be
needed to constrain the amplitude of the homogenities required to have a
sufficiently high SNR to detect with MIRI a potential modulation at 10 $\mu$m
caused by clouds.

\subsection{Variability associated with CH$_4$ abundance variations}\label{sec:ch4}

\begin{figure*}[btp]
\begin{centering}
\includegraphics[width=1.0\linewidth]{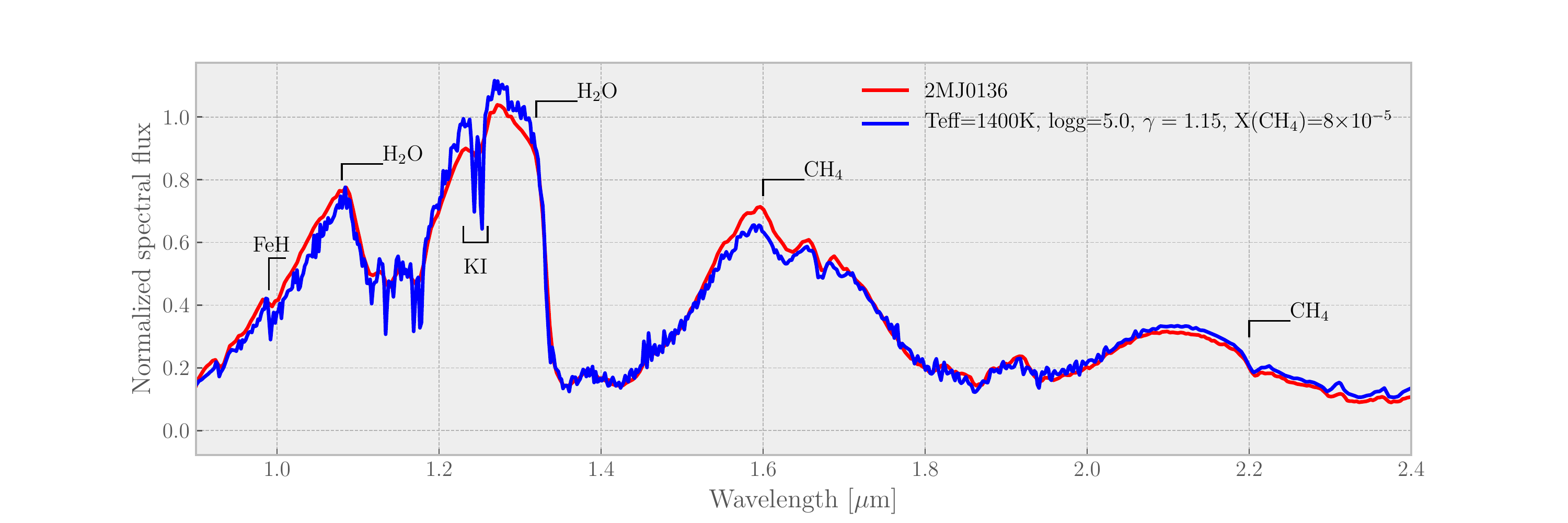}
\caption{Red: SpeX prism spectrum of 2M~0136 \citep{burgasser:2008}. Blue:
  \texttt{ATMO} model indicating the main spectral
  signatures: FeH, H$_2$O, KI lines, and CH$_4$ at 1.6 and 2.2
  $\mu$m.} \label{fig:spec_2M0136}
\end{centering}
\end{figure*}

\begin{figure*}[btp]
\begin{centering}
\includegraphics[width=1.0\linewidth]{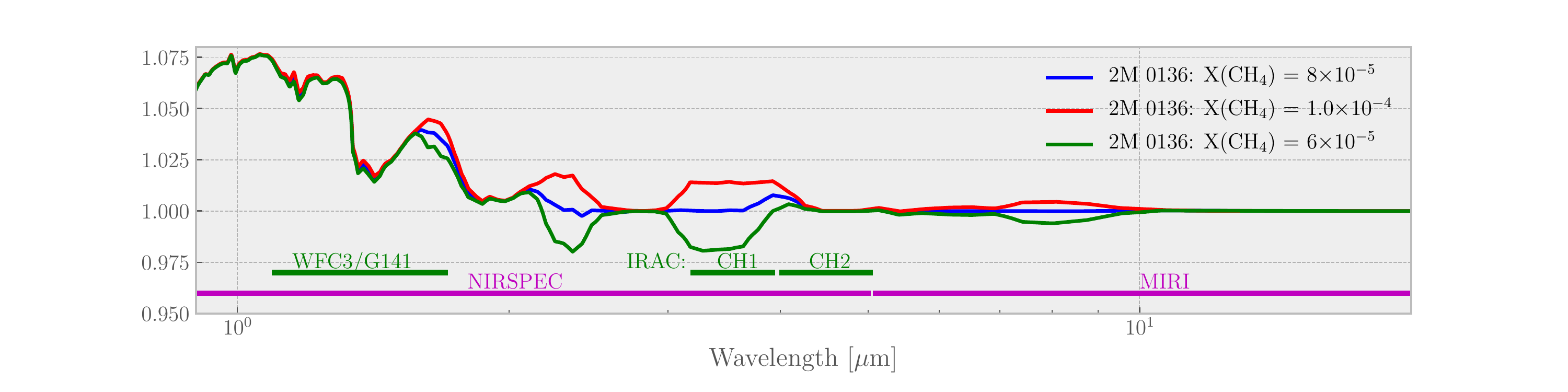}
\caption{Bright-to-faint spectral ratio of 2M 0136 between 0.9 and 20 $\mu$m with a reduced and
  increased abundance of CH$_4$ ($\pm$ 25 \% of the background abundance).} \label{fig:ch4}
\end{centering}
\end{figure*}

In Sect.~\ref{fig:2M1821-2M0136} and \ref{fig:pso}, we have modeled
out-of-equilibrium chemistry by keeping the CH$_4$ abundance constant. This is
justified for 2M~1821 and PSO~318.5-22 since these dwarfs are in the L-dwarf
regime: CH$_4$ is not sufficiently abundant to impact significatively the
spectrum, hence there is no modulations associated to CH$_4$ abundance
fluctuations to expect. This is not the case for 2M~0136 which is in the T-dwarf
regime. Figure~\ref{fig:spec_2M0136} shows the observed SpeX prism spectrum of
2M~0136 and the \texttt{ATMO} model that we have used for the background
state. In addition to FeH, H$_2$O, and the KI lines, CH$_4$ spectral signatures are
clearly visible at 1.6 and 2.2 $\mu$m. We have constrained an abundance of
8$\times$10$^{-5}$ (mole fraction) to reproduce the observed spectrum. The
equilibrium abundance 
profile of CH$_4$ reaches such an abundance at a pressure level of 1.5 bar. We
have estimated a difference of $\le$ 1~\% in the CH$_4$ mole fraction
at this level induced by the modulation. This is mainly because the quenching pressure is higher
than the regions in which we change the temperature (top pressure is at
3.7 bar see Tab.~\ref{tab:2M1821-2M0136}). Under 
the assumption of vertical quenching of CO/CH$_4$ chemistry, we would expect the
same level of difference in the quenched part of the CH$_4$ abundance
profile. By varying the CH$_4$ abundance at 1~\% with respect to the background
quenched value, we have observed negligible differences
in the spectral modulation. We have therefore computed models with a much larger
fluctuation with a reduced or increased CH$_4$ abundance of 25 \% to
explore the impact of CH$_4$ on spectral modulations in the JWST spectral
coverage. The resulting spectral ratio are
shown in Fig.~\ref{fig:ch4}. This large-amplitude (25\%-level) abundance
modulation has a clearly visible impact at 1.6-1.7, 2.0-2.5, 3.0-4.0, and 7.0-9.5
$\mu$m. A 25\%-level modulation in CH$_4$ abundance can
 be expected if we extend the modulation of the temperature gradient to
 higher up in the atmosphere, such a level is obtained for
 $P_\mathrm{mod,min}=1.5$~bar for the 2M~0136 model.
 Such modulations of the temperature gradient in the region around one bar could,
 therefore, explain the possible CH$_4$ variability detection in the spectrum of
 the very red L-dwarf companion VHS J1256-1257b obtained by \citet{zhou:2020}
  and could be further constrained with future JWST observations.

\section{Conclusions and discussion} \label{sec:conclusion}

We have shown in this paper that:
\begin{itemize}
\item The evolution of the HST spectral-flux ratio  between different objects
  along the cooling 
  sequence of brown dwarfs from late-L to early-T has similar characteristics to
  the spectral rotational modulations in 
  individual objects at the L/T transition. This suggests that the main
  parameter that varies during the cooling sequence, i.e. temperature, might
  play a significant role in the rotational modulations of objects at the L/T
  transition.
\item By varying the temperature gradient in the structure of T and L dwarfs cloudless
  models, we are indeed able to reproduce the presence or absence of increased
  variability amplitude 
within the 1.45 um water absorption feature in the HST spectral modulation.
\item By changing the temperature gradient in the atmosphere, we are also able
  to reproduce the nearly anti-correlated modulation between HST and Spitzer
  data for PSO~318.5-22. This signature fits with our interpretation that a
  modification of the temperature gradient is responsible for the L/T transition
  and could be the natural consequence of CO/CH$_4$ radiative convection.
\item The impact of CH$_4$ abundance variations is negligible for the spectral
  modulations of the objects studied in this
  paper (around 1~\% abundance variations), however, temperature variations
  higher up in the atmosphere could lead to
  stronger variability associated to such CH$_4$ abundance variations.
\end{itemize}

Future spectral-modulation studies with the large spectral coverage of, e.g., JWST on
early and late T dwarfs might help identifying spectral modulations associated
to CH$_4$. For early T dwarfs, CH$_4$ abundance fluctuations can impact a large
part of the atmosphere
because of out-of-equilibrium chemistry and these signatures could be
interesting to probe vertical and horizontal quenching induced by winds and
convection.

We show here that the representative pressure-temperature modulations we
introduced in the modeled atmospheres can lead to near-anti-correlated phase shifts
between 1.1-1.7 $\mu$m and 3-5 $\mu$m rotational lightcurves, similar to the
phase shifts observed for the L7 dwarf PS~318.5-22 \citep{biller:2018} and for the
T6.5 dwarf 2M2228 \citep{buenzli:2012}.
 An anti-correlation
could be the sign of a change of the temperature gradient in the atmosphere,
possibly also at the origin of the L/T transition and induced by CO/CH$_4$
radiative convection. Such modifications of the PT profile could also be at the
origin of the unexpected behavior of the KI-lines modulation in Luhman 16B, as
suggested by \citet{kellogg:2017}. In general, modulations at the resolution of
spectral lines might be much easier to produce with temperature changes than
cloud (continuum) opacity changes.

 The $\sim$200 degree phase shift between the 1.1-1.7 $\mu$m and 3-5 $\mu$m
 lightcurves observed for PSO~318.5-22 \citep{biller:2018} may, however, not be
 typical. Although \citet{buenzli:2012} identified a similar magnitude phase
 shift in the lightcurves of the T6.5-type 2M2228, the five simultaneous
 lightcurves in that source showed pressure-dependent offsets \citep[also
   confirmed in][]{yang:2016}. In addition, \citet{yang:2016} founds smaller
 phase shifts 
 (10-25 degrees) in late L dwarfs and L/T transition dwarfs, demonstrating that
 different brown dwarfs likely have different vertical-longitudinal atmospheric
 structures.  If we explain an exact
anti-correlation with a change in the temperature gradient in a region of the
surface of the brown dwarf, the $\sim$20$^\circ$ lag from the exact
anti-correlation for PSO~318.5-22 could be the sign of wind advection of this region
probed at different depth by HST and Spitzer data (hence at different radiative
timescales). This shift could be similar 
to the shift of the hot-spot at different bandpasses caused by Hot Jupiters
atmospheric circulation and is of 
the order of what is observed in the Hot Jupiter regime. The shape of the
``spot'' may be, however, more complex since high-temperature spots qualitatively
similar to the Great Red Spot in Jupiter atmosphere have been excluded by the
lightcurve evolution data \citep[see][]{apai:2017}.
The wind speed
 for brown dwarfs \citep[e.g. 650$\pm$310 m/s][]{allers:2020} may be relatively
smaller than the zonal jet in Hot Jupiters \citep[e.g. 3-4
  km/s][]{showman:2008}. Hence this link is relatively 
speculative at this stage and further studies are needed to see if the
atmospheric circulation in brown dwarfs could lead to such phase shifts at
different bandpasses.

In
general, we have mainly focused our study on reproducing the spectral
characteristic of the bright-to-faint spectral ratios of the modulation. A
future line of work is indeed to look further into multi-dimensional
time-dependent simulations of diabatic convection and also study if they can
reproduce the typical lifetime of the features responsible of the rotational
modulations and the lightcurve time evolution \citep[typically 5-40
  hours][]{apai:2017}. However we point out that if a switch at the global
scale between diabatic and adiabatic convection is responsible for the L/T
transition when CO is converted into CH$_4$, it is likely going to produce
large-scale inhomogeneities in the form of waves and large-scale circulation
patterns. Assuming large-scale features with sizes of the 
order of half the radius of Jupiter (assuming a $\sim$ 10~\% cover fraction of a
brown dwarf visible disk) and typical convective velocities between
10$^4$ and 10$^5$ cm/s \citep{atkinson:1997}, the typical order of magnitude of
the lifetime of these 
large-scale features would be between 10 hours and 4 days which is fairly
consistent with the observed lifetime.
As a consequence, a switch between diabatic and
adiabatic convection could indeed explain relatively well the time evolution of
the spectra of brown dwarfs at the L/T transition. The magnitude of the
temperature fluctuations expected for standard convection is relatively modest
 \citep[typically a few percent, e.g.][]{aurnou:2008}
and is mainly a function of the level of 
super-adiabaticity. This could be much larger for diabatic
convection that can get a larger temperature fluctuation associated to a
compositional fluctuation for the same level of ``super-diabacity'' (defined as
the difference of the unstable diabatic potential temperature gradient to the
stable potential temperature gradient). Numerical simulations are required to
study in more details the magnitude, lengthscale and timescale of the
temperature fluctuations associated to diabatic convection.

This paper demonstrates that 
some key observed spectral charateristics of rotational modulations, based on
low-resolution time-resolved NIR spectrophotometry, can be
interpreted in term of temperature variations. This result suggests that modulations
from temperature variations and cloud-opacity variations are
degenerate.
We are not arguing that no modulations is coming from clouds: if
present and not homogeneous they are likely to cause rotational
modulations \citep[see e.g.][for 1D time-dependent
  models]{tan:2019}. Furthermore, clouds can induce 
temperature fluctuations because of 
radiative heating and cooling. However these modulations on their owns are not a
particularly 
strong sign of the presence of clouds especially in a convective atmosphere with
complex chemistry (e.g. CO/CH$_4$ radiative convection).
The detection of direct cloud spectral signatures, e.g. the silicate absorption feature at
10 $\mu$m, would help to confirm that the observed variability can be driven by
clouds although it will not necessarly exclude temperature variations or other
mechanisms to be at play. Future
studies (e.g. with JWST) looking at the 
differences in the rotational spectral modulation of objects with and without
the silicate absorption feature may give us some insight on how to distinguish
cloud-opacity fluctuations from temperature fluctuations.

\begin{acknowledgements}
PT and AE acknowledges supports by the European Research Council under Grant
Agreement ATMO 757858. PT thanks S. A. Metchev for useful discussions about the
observations of spectral modulations at the Exoclime conference. This work is
also partly supported by the ERC grant 787361-COBOM. 
\end{acknowledgements}

\bibliographystyle{aa}
\bibliography{main}


\end{document}